\documentclass[pra,aps,amsfonts,amsmath,superscriptaddress,%
  twocolumn,showpacs,floatfix]{revtex4}
\usepackage{amsfonts}
\usepackage{amsmath}
\usepackage{bm}
\usepackage{epsfig}
\newcommand{\Ref}[1]{Ref.\@ \cite{#1}}
\newcommand{\Refs}[1]{Refs.\@ \cite{#1}}
\newcommand{\Eq}[1]{Eq.\@ (\ref{#1})}

\newcommand{\Fig}[1]{Fig.\@ \ref{#1}}
\newcommand{\nablav}{\bm{\nabla}}
\newcommand{\pv}{\bm{p}}
\newcommand{\rv}{\bm{r}}

\newcommand{\lp}{l^\prime}
\newcommand{\np}{n^\prime}

\renewcommand{\Im}{\mathrm{Im}\,}

\begin{document}
\title{Microscopic description of the twist mode\\
in normal and superfluid trapped Fermi gases}
\author{Marcella Grasso}
\affiliation{Dipartimento di Fisica e Astronomia and INFN, Via Santa
  Sofia 64, I-95123 Catania, Italy}
\affiliation{Institut de Physique Nucl\'eaire, 15 rue Georges
  Cl\'emenceau, F-91406 Orsay Cedex, France}
\author{Michael Urban}
\affiliation{Institut de Physique Nucl\'eaire, 15 rue Georges
  Cl\'emenceau, F-91406 Orsay Cedex, France}
\author{Xavier Vi\~nas}
\affiliation{Departament d'Estructura i Constituents de la Mat\`eria,
Facultat de F\`{\i}sica, Universitat de Barcelona, Diagonal 647,
  E-08028 Barcelona, Spain}
\begin{abstract}
We investigate the ``twist'' mode (rotation of the upper against the
lower hemisphere) of a dilute atomic Fermi gas in a spherical
trap. The normal and superfluid phases are considered. The linear
response to this external perturbation is calculated within the
microscopic Hartree-Fock-Bogoliubov approach. In the normal phase the
excitation spectrum is concentrated in a rather narrow peak very close
to the trapping frequency. In the superfluid phase the strength starts
to be damped and fragmented and the collectivity of the mode is
progressively lost when the temperature decreases. In the weak-pairing
regime some reminiscence of the collective motion still exists,
whereas in the strong-pairing regime the twist mode is completely
washed out. The disappearance of the twist mode in the strong-pairing
regime with decreasing temperature is interpreted in the framework of
the two-fluid model.
\end{abstract}
\maketitle
\section{Introduction}
The experimental and theoretical development of Bose-Einstein
condensation of trapped bosonic atoms \cite{Anderson} has also
triggered the investigation of trapped atomic Fermi gases at very low
temperature \cite{DeMarco}. One of the main goals in the research of
these Fermi systems is to detect the transition from the normal to the
superfluid phase, associated with the appearance of a macroscopic
order parameter of strongly correlated Cooper pairs below a certain
critical temperature $T_c$. In order to have an attractive $s$-wave
interaction which can provide the pairing correlations, the atoms must
be trapped and cooled in two different hyperfine states as has been
achieved in several recent experiments (see, e.g., \Ref{OHara}). From
theoretical side, the pairing problem of trapped fermions has been
studied from different points of view
\cite{Houbiers,BaranovPetrov1,BruunCastin}.

Besides the ground-state properties, there is also interest in knowing
the spectrum of collective excitations. As we stated above, ultracold
atomic Fermi gases are assumed to become superfluid below $T_c$, and
it is therefore important to study low energy collective modes also in
the superfluid phase \cite{BaranovPetrov2,BruunMottelson}. Having
different properties in the normal and superfluid phases, these
excitations can serve as experimental signals for superfluidity. For
instance, the frequencies of breathing modes of trapped atomic Fermi
gases measured in recent experiments \cite{Kinast} give strong
indications that the superfluid phase has been reached.

It is interesting to compare the situation of trapped fermionic atoms
to that of atomic nuclei, which can also show a superfluid behavior.
Contrary to the nuclear case, the fact that the interaction in atomic
gases is tunable experimentally allows to study the collective modes
in different regimes. For dilute systems, the atom-atom interaction
can be parametrized by a zero-range force proportional to the $s$-wave
scattering length between atoms in two hyperfine states
\cite{BruunCastin,BruunMottelson}. By changing the applied magnetic
field around a Feshbach resonance \cite{Roberts}, the $s$-wave
scattering length can be modified. In one limiting regime, that of
weak pairing, which is similar to the situation in atomic nuclei, the
pairing results only in a small perturbation to the response of the
system to the external probe. In the other limit of strong pairing the
response is dominated by the effects of superfluidity.

Many of the collective excitations show features proper to Landau's
zero sound modes in bulk Fermi liquids \cite{BaymPethick} which for
finite Fermi systems translate into modes analogous to those of an
elastic body \cite{Lamb}. Since the trapped atomic Fermi gases contain
a very large number of atoms, the single-particle orbital angular
momenta near the Fermi surface can also become very
large. Consequently, important orbital effects such as excitations
having angular momentum and parity $J^P = 1^+$ and $2^-$ will exist,
which correspond to magnetic resonances of M1 or M2 type,
respectively, in atomic nuclei. The $2^-$ excitation is the so-called
twist mode, in analogy to the quadrupole torsional vibration of an
elastic sphere \cite{Lamb,Holzwarth}. From a macroscopic point of
view, the twist consists of a coherent counterrotation of the
particles in the upper hemisphere against those in the lower
hemisphere. For small amplitudes, it corresponds to a purely kinetic
excitation without spatial distortion of the equilibrium shape.

The twist mode has been studied in different Fermi systems. In nuclei,
this mode has been analyzed from a semiclassical point of view within
a fluid-dynamical description \cite{Holzwarth}. From a quantum
mechanical point of view, this mode has been studied so far only for
magic nuclei (i.e., without pairing) such as $^{90}$Zr and $^{208}$Pb
\cite{Schwesinger,Ponomarev}. More recently, some experimental effort
has been done to detect this mode by backward inelastic electron
scattering \cite{Ca48Zr90}. A direct evidence for the existence of the
orbital twist mode (to be distinguished from the $2^-$ spin-flip mode)
in nuclei has been achieved by comparing electron and proton
scattering cross-sections of $^{58}$Ni \cite{Ni58}. The twist mode has
also been theoretically studied in metallic clusters \cite{Nesterenko}
although it has not yet been detected.

So far, the theoretical study of the twist mode in trapped atomic
Fermi gases has been done in the hydrodynamical description and in the
normal phase only \cite{VinasRoth}. In the case of a $s$-wave
interaction, a moderate shift of the twist frequency of about $10\%$
with respect to the non-interacting case was found, which is
consistent with the fact that for a transverse zero-sound the $s$-wave
interaction does not contribute to the restoring force
\cite{Schwesinger,SchwesingerPingel}.

In the present article, our aim is different. We will analyze the
effect of pairing correlations on the twist mode. This effect has not
been considered in any of the theoretical studies mentioned above,
neither for atomic nuclei nor for metallic clusters or trapped Fermi
gases. Of particular interest can be the study of the strong pairing
regime, because it is known that in this case the low-energy
collective modes are strongly affected by the pairing and can become
signatures that the superfluid phase is reached
\cite{BaranovPetrov2,BruunMottelson}.

The paper is organized as follows. In Sec.\@ 2, we sketch the
derivation of the twist response function in the superfluid phase,
using a Hartree-Fock-Bogoliubov or Bogoliubov-de Gennes framework
\cite{RingSchuck,deGennes}. In Sec.\@ 3, we consider the twist mode
in the normal phase within a quantum-mechanical description. Sec.\@ 4
is devoted to the study of the twist mode in the superfluid phase in
the cases of weak and strong pairing correlations. Finally, our
conclusions are laid in the last section.

\section{Quasiparticle response function}
In this article we will consider an atomic Fermi gas (atomic mass $m$),
trapped in a spherical harmonic trap with frequency $\Omega$. We
assume that the atoms equally occupy two hyperfine states, denoted by
$\sigma=\pm1$. Because of the low density of the gas, the interaction
between the atoms can be regarded as pointlike and its strength can be
parametrized by the s-wave atom-atom scattering length $a$. In order
to simplify the notation, we will express all quantities in harmonic
oscillator (h.o.) units, i.e., frequencies in units of $\Omega$,
energies in units of $\hbar \Omega$, temperatures in units of $\hbar
\Omega / k_B$, and lengths in units of the oscillator length $l_{ho} =
\sqrt{\hbar / (m \Omega)}$. Furthermore, instead of the scattering
length we will use the coupling constant $g = 4 \pi a / l_{ho}$ as
parameter of the interaction strength.

The twist is a motion where the upper and lower hemispheres rotate in
the opposite sense back and forth around the $z$ axis with an angle
proportional to $z$. This mode can be excited in both spherical and
deformed (with a rotation axis) systems. Such a motion can be
generated by the operator $z L_z$, where $L_z = -i (x \nabla_y - y
\nabla_x)$ denotes the $z$ component of the angular momentum
operator. Restricting our description to small amplitudes, we can use
linear response theory in order to treat the oscillations around
equilibrium. Then the main problem consists in calculating the
equilibrium state. In order to describe the system in the superfluid
phase, this is done within the framework of a Hartree-Fock-Bogoliubov
(HFB) or Bogoliubov-de Gennes \cite{RingSchuck,deGennes} calculation
similar to that presented in \Ref{BruunCastin}, but with the modified
regularization scheme for the gap equation described in
\Refs{Bulgac,Grasso}. We refer to \Ref{Grasso} for more details about
our approach. The calculation provides the wave functions
$u_\alpha(\rv)$ and $v_\alpha(\rv)$ satisfying the HFB equations
\begin{equation}
\begin{split}
[H_0 + W(\rv)] u_{\alpha}(\rv) + \Delta(\rv) v_\alpha(\rv) 
  &= E_\alpha u_\alpha(\rv)\,,\\
\Delta(\rv) u_\alpha(\rv) - [H_0 + W(\rv)]  v_\alpha(\rv) 
  &= E_\alpha v_\alpha(\rv) \,.
\end{split}
\label{hfbeq}
\end{equation}
Here $H_0$ denotes the hamiltonian of the non-interacting h.o. minus
the chemical potential, $H_0 = (-\nablav^2+r^2)/2-\mu$, while the
interaction is accounted for in a self-consistent way through the
Hartree potential $W(\rv)$ and the pairing field $\Delta(\rv)$.

Now let us consider the retarded correlation function
\begin{equation}
\Pi_0(\omega) = -i\int_0^\infty dt\,e^{i\omega t}
  \langle\langle[Q(t),Q(0)]\rangle\rangle\,,
\label{defpi0}
\end{equation}
where $\langle\langle\cdot\rangle\rangle$ means the thermal
average. In our case, $Q$ is the twist operator
\begin{equation}
Q(t) = \sum_{\sigma = \pm 1}\int d^3 r\psi^\dagger_\sigma(t,\rv) z L_z
  \psi_\sigma(t,\rv)\,.
\end{equation}
The field operator $\psi$ can be expressed in terms of
quasiparticle creation and annihilation operators $b^\dagger$ and $b$
as follows:
\begin{multline}
\psi_\sigma(t,\rv) = \sum_{nlm}\Big(b_{nlm\sigma}u_{nlm}(\rv)e^{iE_{nl}t}\\
  -\sigma b^\dagger_{nlm-\sigma}v^*_{nlm}(\rv)e^{-iE_{nl}t}\Big)\,.
\end{multline}
Separating the radial and angular dependence of the wave functions,
$u_{nlm}(\rv) = u_{nl}(r)Y_{lm}(\theta,\phi)$ and $v_{nlm}(\rv) =
v_{nl}(r)Y_{lm}(\theta,\phi)$, one obtains after a straight-forward
but tedious calculation the following result:
\begin{widetext}
\begin{multline}
\Pi_0(\omega) = 2\sum_{n\np l\lp m}
  m^2\Big|\int d\Omega Y^*_{lm}(\theta,\phi)\, \cos\theta\,
  Y_{\lp m}(\theta,\phi)\Big|^2\\
\times \Big[
  \frac{(E_{\np\lp}+E_{nl})[1-f(E_{nl})-f(E_{\np\lp})]}
          {(\omega+i\eta)^2-(E_{\np\lp}+E_{nl})^2}
  \Big(\int_0^\infty dr\,r^3
    [u_{nl}(r)v_{\np\lp}(r)-v_{nl}(r)u_{\np\lp}(r)]\Big)^2\\
  +\frac{(E_{\np\lp}-E_{nl})[f(E_{nl})-f(E_{\np\lp})]}
          {(\omega+i\eta)^2-(E_{\np\lp}-E_{nl})^2}
  \Big(\int_0^\infty dr\,r^3
    [u_{nl}(r)u_{\np\lp}(r)+v_{nl}(r)v_{\np\lp}(r)]\Big)^2\Big]\,.
\label{pi0}
\end{multline}
\end{widetext}
In deriving this formula, we have used the anticommutation relations
between the operators $b$ and $b^\dagger$ [$\{b_\alpha, b_\beta\} =
\{b^\dagger_\alpha, b^\dagger_\beta\} = 0$, $\{b_\alpha,
b^\dagger_\beta\} = \delta_{\alpha\beta}$] as well as the relation
$\langle\langle b^\dagger_\alpha b_\beta \rangle\rangle = f(E_\alpha)
\delta_{\alpha\beta}$, where $f$ denotes the Fermi distribution
function, $f(E) = 1/(e^{E/T}+1)$. Note that the relative signs
appearing in the radial integrals in \Eq{pi0} are different from those
obtained, e.g., for the case when $Q$ is a multipole operator as in
\Ref{BruunMottelson}. The reason is that the twist operator is odd
under time reversal, i.e., $\int d^3r f^*(\rv) zL_z g(\rv) = -[\int
d^3r g^*(\rv) zL_z f(\rv)]^*$. The angular matrix element in \Eq{pi0}
can be computed explicitly, with the simple result
\begin{multline}
\sum_m m^2\Big|\int d\Omega Y^*_{lm}(\theta,\phi)\, \cos\theta\,
  Y_{\lp m}(\theta,\phi)\Big|^2\\
= \begin{cases}
  \frac{(\lp-1)\lp(\lp+1)}{15}&\mbox{if $\lp = l+1$\,,}\\
  \frac{(l-1)l(l+1)}{15}&\mbox{if $l = \lp+1$\,,}\\
  0&\mbox{otherwise}\,.
\end{cases}
\end{multline}
Therefore the numerical task of calculating $\Pi_0$ reduces
essentially to calculating the radial integrals and the triple sum
over $n$, $\np$, and $l$.

In general it is not sufficient to calculate the free quasiparticle
response $\Pi_0$. Rather one has to calculate the QRPA response, which
accounts for correlations with the quantum numbers corresponding to
the excitation under consideration in the ground state. However,
because of the particular form of the interaction used here, it is
clear that there cannot be any ground state correlations with the
quantum numbers of the twist mode ($J^P = 2^-$). Therefore the QRPA
response function, $\Pi$, is just equal to the free quasiparticle
response function, $\Pi_0$ \cite{Nesterenko}. In this sense the
situation for trapped atoms is different from that in nuclei, where
the spin-orbit part of the interaction leads to a (small) change of
the twist response function \cite{Schwesinger}, e.g., through the
coupling between the twist mode and the spin-flip mode, which is
excited by the operator $(\rv\otimes\bm{\sigma})_{20}$.

In the remaining part of this article we will show numerical results
for the strength function $S(\omega) = -\Im \Pi(\omega)/\pi$ which we
calculate from \Eq{pi0} with a finite width $\eta$ for each peak.

\section{Normal phase}
Let us first look at the normal phase of the system at zero
temperature, i.e., we artificially put $\Delta = 0$ in \Eq{hfbeq}. For
this situation, there exist microscopic descriptions of the twist mode
in nuclei \cite{Schwesinger,Ponomarev} and in metal clusters
\cite{Nesterenko}. However, for the twist mode in trapped atomic gases
there exists only a calculation \cite{VinasRoth} following the fluid
dynamical approach developped by Holzwarth for the nuclear case
\cite{Holzwarth}. This fluid dynamical approach allows to predict the
twist frequency, but it cannot answer the question if the twist mode
as a collective motion exists at all \cite{Holzwarth}.

In the case of a non-interacting h.o., it is straight-forward to show
that the operator $z L_z$ excites only transitions with $\omega = 1$
(in units of $\hbar\Omega$). Therefore, in the non-interacting h.o.,
the total strength is concentrated at $\omega = 1$. If now the Hartree
potential $W$ is switched on, two effects are to be expected:

a) The energy difference between neighboring shells becomes larger
 (smaller) in the case of an attractive (repulsive) interaction.
 Therefore, the twist frequency will be shifted upwards (downwards).
 This effect has been described quantitatively within the fluid
 dynamical approach \cite{VinasRoth}.

b) The degeneracy of states with different $l$ is lifted, and we
 therefore expect a fragmentation of the strength of the twist mode
 into many particle-hole states corresponding to transitions $n,l\to
 n,l+1$ and $n,l\to n+1,l-1$ [remember that for given quantum numbers
 $n$ and $l$, the number of h.o. quanta is $2(n-1)+l$].

\begin{figure}
\begin{center}
\epsfig{file=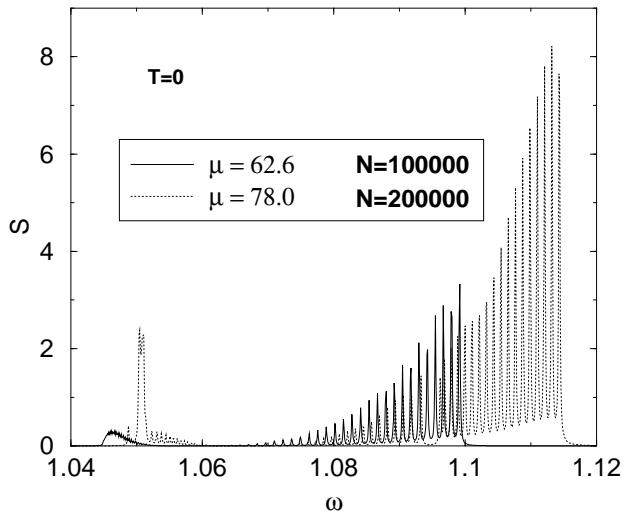,width=8.2cm}
\end{center}
\caption{Strength function $S(\omega)$ ($\times 10^9$; in h.o. units)
for the twist mode in a gas of $10^5$ (solid line) and $2\times 10^5$
(dashed line) trapped $^6$Li atoms at zero temperature without pairing
($\Delta = 0$) as function of the frequency $\omega$ (in units of
$\hbar\Omega$).}
\label{fig1}
\end{figure}

Both effects can be observed in \Fig{fig1}, where we display the
strength function $S(\omega)$ of the twist mode as a function of the
excitation energy for two systems of $^6$Li with different numbers of
atoms (scattering length $a = -2160 a_0$ \cite{Abraham}, where $a_0$ is
the Bohr radius) in a trap with a frequency of $\Omega = 2\pi\times
144$ Hz, corresponding to a coupling constant $g = -0.4$ in
h.o. units. In order to show the fragmentation of the mode, we display
the response function in a small energy interval (containing $100\%$
of the total strength) with a very high resolution ($\eta = 10^{-4}$).
Let us first look at the result corresponding to $10^5$ particles in
the trap ($5\times 10^4$ particles per spin state, chemical potential
$\mu = 62.6$). One can clearly see that the average frequency is
higher than $1$ and that the strength is fragmented into two series of
peaks, corresponding to the two series of transitions mentioned above
under b). With $2\times 10^5$ particles ($10^5$ particles per spin
state, $\mu = 78.0$), the Hartree field is stronger and therefore both
effects, fragmentation and shift of the average frequency, are
enhanced.

The difference shown by the strength at low frequencies of the systems
containing $10^5$ and $2\times 10^5$ particles is related to the
different single-particle spectra of these two systems. In the case of
$10^5$ particles, the Hartree field $W$ breaks the accidental
degeneracy of the non-interacting h.o. single-particle levels, but the
different h.o. major shells are still separated. However, when the
number of particles in the trap grows, the Hartree field becomes
strong enough to mix different h.o. major shells. This leads more or
less accidentally to the fact that in the case of $2\times 10^5$
particles the transition energies of the series $n,l\to n+1,l-1$ with
$2(n-1)+l = 82$ (the major shell number 82 is the last one lying
completely below the Fermi level) are almost degenerate at
$\omega\approx 1.05$.

In order to compare our results quantitatively with the predictions
obtained within the fluid-dynamical approach, we define an average
frequency according to
\begin{equation}
\omega_\mathit{av} = \frac{\int_0^\infty d\omega \omega S(\omega)}
  {\int_0^\infty d\omega S(\omega)}\,.
\label{omegaav}
\end{equation}
For both cases considered here, this average frequency is in perfect
agreement with the frequency $\omega_\mathit{fd}$ predicted in
\Ref{VinasRoth} in the framework of the fluid-dynamical approach: for
$N = 10^5$ atoms, $\omega_\mathit{av}=1.088$ and $\omega_\mathit{fd} =
1.087$, and for $N = 2\times 10^5$ atoms, $\omega_\mathit{av}=1.100$
and $\omega_\mathit{fd} = 1.101$. It should also be emphasized that
the width of the interval over which the strength is distributed is
very narrow compared with the average frequency of the twist mode. It
is therefore justified to speak about a collective excitation.

\section{Superfluid phase}

Let us now consider the superfluid case. It is interesting to analyse
how the properties of the twist mode are modified when pairing
correlations are taken into account in the calculations and the full
HFB equations are solved. We will show that the structure of the
strength function and the collectivity of the twist mode are strongly
affected by pairing correlations and we will study this dependence at
different temperatures for two systems with different numbers of
atoms.
 
We set the coupling constant $g$ equal to $-1$ in h.o. units. For
$^6$Li atoms with a scattering length $a=-2160 a_0$ this corresponds
to a trapping frequency of $\omega=2\pi \times 817$ Hz. (We chose a
stronger coupling than in the previous section in order to be able to
study the case of strong pairing, which would be possible only for
extremely large numbers of particles if $g = - 0.4$.)  We shall
consider two cases for the trapped gas: a) A small system with around
1800 atoms (weak pairing regime, $\Delta < \hbar \Omega$); b) A big
system with around $3.6\times 10^4$ atoms (strong pairing regime,
$\Delta > \hbar \Omega$). For both cases we will take into account
different temperatures and analyze how the twist mode evolves when the
critical temperature $T_c$ of the phase transition is approached and
crossed.

Before passing to consider the two cases we would like to mention that
in atomic nuclei, which are the only systems for which the twist mode
has been observed so far, one is always in the weak pairing regime,
the relation $\Delta < \hbar \Omega$ being always satisfied.

\begin{figure}
\begin{center}
\epsfig{file=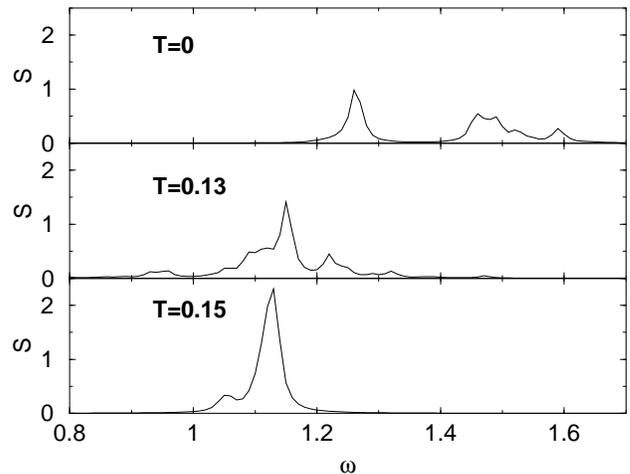,width=8.2cm}
\end{center}
\caption{Strength function $S(\omega)$ ($\times 10^5$; in h.o. units)
for the twist mode in a system with about 1800 atoms of $^6$Li at
$T=0$ (top), $T=0.13$ (middle) and $T=0.15$ (bottom; $\omega$ and
$T$ in units of $\hbar \Omega$) \label{fig2}}
\end{figure}

a) Weak pairing regime: The chemical potential $\mu$ in this case is
chosen equal to 16. We show in \Fig{fig2} the strength function for
three values of the temperature: $T = 0$ (top), $T = 0.13$ (middle)
and $T = 0.15$ (bottom). The three cases correspond to values of the
pairing field in the center of the trap of $\Delta (r=0) = 0.67$, 0.26
and 0, respectively (in h.o. units).  In the last case (bottom of the
figure) the gas is in the normal phase: we observe that in the normal
phase the strength function is concentrated at about $\omega = 1.12$
(this is slightly higher than in \Fig{fig1} because of the stronger
coupling, which leads to a stronger Hartree field). If we lower the
temperature, the superfluid transition takes place; the effect on the
strength function is to push its structure towards higher values of
the energy. Qualitatively this can be understood by replacing the
single-particle energies $\epsilon_{nl}$ by the quasiparticle energies
$E_{nl}\approx \sqrt{(\epsilon_{nl}-\mu)^2+\Delta^2}$, where $\Delta$
denotes the average matrix element of the pairing field at the Fermi
surface. Neglecting the effect of the Hartree field for the moment,
one obtains in this way a shift of the twist frequency from $1$ to a
higher value which lies between $\sqrt{1+4\Delta^2}$ and
$\Delta+\sqrt{1+\Delta^2}$. To see this, let us consider two limiting
cases: If the chemical potential lies exactly on a single-particle
level (half-filled shell), $\mu = N_F+3/2$, a transition of the type
$N_F\to N_F+1$, for example, corresponds to the creation of two
quasiparticles with energies $E_{N_F} = \Delta$ and $E_{N_F+1} =
\sqrt{1+\Delta^2}$. In the other limiting case, the chemical potential
lies between two single-particle levels (closed shell), $\mu = N_F+2$,
and the twist mode corresponds to the excitation of two quasiparticles
having each the energy $E_{N_F}$ = $E_{N_F+1} = \sqrt{1/4+\Delta^2}$.

Moreover, as one can also observe in \Fig{fig2}, the excitation mode
becomes less collective and, due to pairing, more and more damped and
fragmented if one goes from $T = 0.13$ to $T = 0$. In the latter case
pairing correlations are more intense and the loss of collectivity and
the Landau damping are consequently more important. A similar Landau
damping effect due to superfluidity has been found in
\Ref{BruunMottelson} for the spin-dipole mode in the weak pairing
regime.

An other interesting effect to notice is the strength below $\omega =
1$ which appears below $T_c$ but disappears at $T = 0$. Obviously this
effect is due to the second term in \Eq{pi0}, which is equal to zero
at $T = 0$.

\begin{figure}
\begin{center}
\epsfig{file=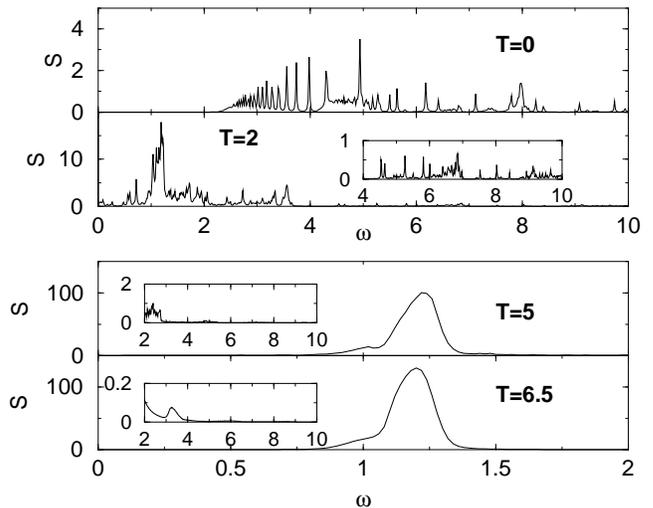,width=8.2cm}
\end{center}
\caption{Strength function $S(\omega)$ ($\times 10^5$; in h.o. units)
in a gas with about $3.6\times 10^4$ atoms at four different
temperatures (from top to bottom): $T=0$, 2, 5, and 6.5 ($\omega$ and
$T$ in units of $\hbar \Omega$).
\label{fig3}}
\end{figure}
\begin{figure}
\begin{center}
\epsfig{file=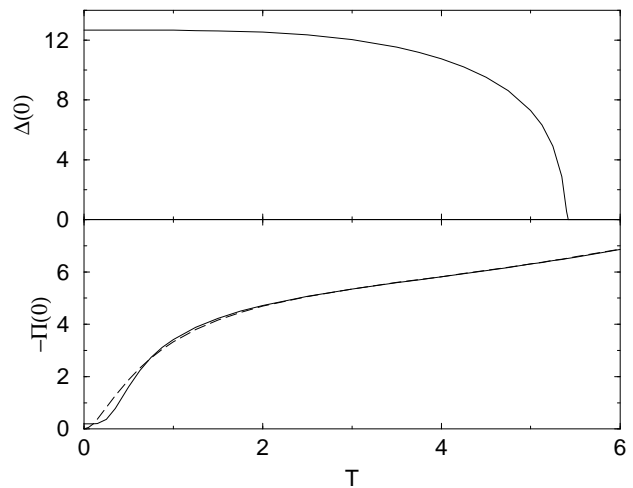,width=8.2cm}
\end{center}
\caption{Temperature dependence of the gap in the center of the trap,
$\Delta(0)$ (top; $\Delta$ and $T$ in units of $\hbar \Omega$), and of
the static response function $-\Pi(0)$ (bottom; $\times 10^{-6}$ in
h.o. units), which is equal to twice the inverse-energy weighted sum
rule, for a gas with $3.6\times 10^4$ atoms. In the lower figure, we
show for comparison the result of the HFB calculation (solid line)
together with the result obtained within the two-fluid model,
\Eq{sumrule} (dashed line).
\label{fig4}}
\end{figure}


b) Strong pairing regime: Let us consider now the case with about
$3.6\times 10^4$ atoms ($\mu = 40$). We present in \Fig{fig3} the
strength function at four temperatures: $T = 0$, 2, 5, and 6.5 (from
top to bottom). In the four cases the central values of the pairing
field are $\Delta (r=0) = 12.7$, 12.5, 9.6, and 0, respectively. In
the latter case (normal phase) we observe a unique peak centered at
about $\omega = 1.2$. Again, the energy is higher with respect to
\Fig{fig1} and with respect to the case a) (\Fig{fig2}) due to the
stronger coupling. A fragmented structure with a very low strength
exists in the energy region from $\omega = 2$ up to $\omega = 4$. When
we lower the temperature, we cross the superfluid transition (see upper
panel of \Fig{fig4}). At $T = 5$ the system is superfluid: we observe
that the main peak still exists, even if the excitation mode is less
collective than in the normal phase case. Also, the fragmented
structure towards $\omega = 2-3$ gets more strength than in the
previous case. The fragmentation becomes much stronger and extends up
to $\omega = 10$ when the temperature is lowered further, as can be
seen in the case $T = 2$. However, the peak at about $\omega = 1.2$ is
still visible. Finally, if we look at the $T = 0$ case, where the
pairing correlations are the strongest, we observe that the main
collective peak completely disappears, while a very fragmented
structure with a low strength remains in the energy region between
$\omega = 2$ and $\omega = 10$. We can thus conclude that at $T = 0$
the collective twist mode does not exist any more. The same conclusion
has been drawn in \Ref{BruunMottelson} for the spin-dipole mode in the
strong pairing regime.

Actually, once the irrotational flow limit (strong pairing) is reached
\cite{Farine,UrbanSchuck}, the superfluid current has an irrotational
velocity field, and the only possible excitations of the superfluid
are density-fluctuation modes. In the language of a two fluid model,
all the other excitations of the gas, such as the twist and the
spin-dipole modes, have to be related to its normal component, as was
discussed in \Refs{Leggett,Betbeder}. When one decreases the
temperature below $T_c$, the number of ''normal'' quasiparticles is
reduced and therefore the strength of the twist mode becomes smaller.
On the other side, the energy spectrum of the normal quasiparticles is
modified, leading to a destruction of coherence between quasiparticles
moving in the same direction \cite{Leggett}. It follows that the mode
is more and more damped when one approaches $T = 0$. Obviously, this
effect will strongly depend on the strength of pairing correlations,
and this is why it is more important in the strong pairing regime.

Let us now discuss the relationship between the strength of the twist
response function and the normal-fluid component of the system in a
more quantitative way. To that end we consider the inverse-energy
weighted sum rule, which is proportional to the real part of the
response function at $\omega = 0$:
\begin{equation}
\int_0^\infty d\omega \frac{S(\omega)}{\omega} = -\frac{1}{2}
\Pi(0)\,.
\end{equation}
Within the two-fluid model it can be shown explicitly (see appendix)
that this quantity is related to the density of the normal-fluid
component of the system, $\rho_n$, by
\begin{equation}
\Pi(0) = -\frac{8\pi}{15} \int_0^\infty d r\, r^6 \rho_n(r)\,.
\label{sumrule}
\end{equation}
In \Fig{fig4} (bottom) we show numerical results for the dependence of
the sum rule on temperature. The solid line represents the full HFB
calculation, while the dashed line corresponds to \Eq{sumrule}. The
agreement is very satisfactory except at extremely low temperature,
where quantum finite-size effects (corrections in
$\hbar\omega/\Delta$, see \Ref{UrbanSchuck}) lead to a non-vanishing
value of the sum rule, whereas the two-fluid model predicts that the
sum rule should go to zero at zero temperature because of the
vanishing normal-fluid component. However, the overall good agreement
confirms our interpretation that only the normal-fluid part of the
system participates in the twist motion. In order to recognize more
easily the regions where the gas is superfluid and normal, and to
observe how pairing correlations decrease by increasing the
temperature, we also plot in \Fig{fig4} (top) the value of the gap at
the center of the trap, $\Delta(0)$. Note that the temperature
dependence of $\Pi(0)$ differs considerably from that of $\Delta(0)$.

To conclude this section, we remark that our approach is only valid in
the regime where collisions between atoms can be neglected. Following
\Ref{BaranovPetrov2}, this means that the mean time between
collisions, $\tau = \rho a^2 v_F (T/\epsilon_F)^2$ (where $\rho$,
$v_F$, and $\epsilon_F$ are the density, Fermi velocity, and Fermi
energy, respectively), must be much larger than the oscillation period
in the trapping potential, $2\pi/\Omega$. Expressed in h.o. units, we
obtain $\Omega\tau/(2\pi) = 6\pi^3/(gT)^2$. In the case of $T = 6.5$
this ratio gives still $4.4$, i.e. an atom performs more than four
oscillations before it collides with another one. Consequently, all
cases we considered are well in the collisionless regime.
\section{Summary and Outlook}
In this article we have studied the twist mode of an atomic Fermi gas
trapped by a spherical harmonic potential in the normal and in the
superfluid phase. The ground state has been obtained by solving the
Bogoliubov-de Gennes equations, using the regularization procedure
introduced in \Ref{Bulgac,Grasso}. The excitations have been treated
within the linear response theory. As the zero-range interaction does
not couple to the twist operator, we analyzed this excitation by
calculating the free quasiparticle response function.

We have analyzed the twist mode without pairing correlations by
setting $\Delta = 0$ in the Bogoliubov-de Gennes equations. We
observed that the strength function is concentrated around an energy
higher than $\omega = 1$. This shift (with respect to the case of a
non-interacting h.o.) is due to the Hartree potential and depends on
the sign of the coupling constant $g$. We have also observed a
fragmentation of the strength which describes the transitions $n, l
\rightarrow n, l+1$ and $n, l \rightarrow n + 1, l - 1$.

In the case of pairing correlations we have shown that the excitation
mode starts loosing its collectivity below the critical temperature
$T_c$. When the temperature is lowered from $T_c$ towards $T = 0$, the
strength function becomes more and more damped and fragmented. In the
weak pairing regime ($\Delta < \hbar \Omega$) this effect is less
pronounced than in the strong pairing regime ($\Delta > \hbar
\Omega$): In the weak pairing case the collective twist mode still
exists at zero temperature. With increasing strength of the pairing
correlations, the collective peak is shifted to higher energies, and
at the same time it becomes more and more broad and fragmented and its
strength decreases. Finally, in the strong pairing limit it completely
disappears at $T = 0$. In fact, it can be predicted that the twist
mode ceases to exist once the pairing is strong enough for the system
to reach its irrotational flow limit \cite{Farine,UrbanSchuck}.

It should be pointed out that, in the normal phase, the twist mode can
only exist in the collisionless regime, since the restoring force for
this collective oscillation comes entirely from the Fermi surface
deformation \cite{Holzwarth}. This means that detecting the twist mode
in the normal phase would be a signal that the system is in the
collisionless regime. This might be of importance since the evidence
for the superfluidity obtained in recent experiments \cite{Kinast}
relies on the assumption that the system is in the collisionless
regime. The subsequent disappearance of the twist mode at lower
temperatures would be a clear signal that the superfluid phase has
been reached. Concerning the possibility to excite the twist mode
experimentally we refer to \Ref{VinasRoth}.

Recently, the twist mode has been measured in open-shell finite nuclei
such as $^{58}$Ni \cite{Ni58}. In the existing theoretical studies of
the twist mode in nuclei pairing correlations have not been taken into
account, i.e., these studies are essentially restricted to
closed-shell (magic) nuclei. Although nuclei are in the weak pairing
regime, we think that a theoretical study of the twist mode in nuclei
taking into account pairing correlations could be very
interesting. Work in this direction is in progress.
\begin{acknowledgments}
One of us (X.V) acknowledges financial support from DGI and FEDER (Spain) 
under grant No. BFM2002-01868 and from DGR (Catalonia) under grant 
No. 2001SGR-00064.
\end{acknowledgments}
\begin{appendix}
\section{Relation between the sum rule and the normal component}
In this appendix we will briefly show how a relationship between
$\Pi(0)$ and the normal-fluid component of a system with strong
pairing ($\Delta\gg 1$ in h.o. units) can be established. A detailed
discussion of some of the topics mentioned here can be found in
\Ref{UrbanSchuck}.

As mentioned in Sec.\@ 4, the inverse-energy weighted sum rule is
proportional to the response $\Pi(0)$ of the system to a static
perturbation with a perturbation hamiltonian $H_1 \propto z L_z$. By
taking the $\hbar\to 0$ limit of the time-dependent HFB equations, one
can derive equations similar to the Vlasov equation for the superfluid
phase (in our case, of course, the time-dependence does not play any
role). The resulting deviation of the Wigner function $\rho(\rv,\pv)$
from its equilibrium value reads
\begin{equation}
\rho_1(\rv,\pv) = \Big(\frac{df(E)}{dE}\Big)_{E=E(\rv,\pv)}
  h_1(\rv,\pv)\,,
\end{equation}
with
\begin{equation}
E(\rv,\pv) = \sqrt{[h(\rv,\pv)]^2+\Delta^2(\rv)}\,,
\end{equation}
where $h(\rv,\pv)$ and $h_1(\rv,\pv)$ denote the Wigner transforms of
$H_0+W(\rv)$ and $H_1$, respectively. Since $\Pi(0)$ is defined as the
expectation value of $z L_z$ in the perturbed system, we can write
\begin{equation}
\Pi(0) = 2 \int\!\frac{d^3 r\, d^3 p}{(2\pi)^3}\,
  \Big(\frac{df(E)}{dE}\Big)_{E=E(\rv,\pv)}\, (zxp_y-zyp_x)^2\,.
\end{equation}
Assuming spherical symmetry and a strongly peaked Fermi surface (i.e.,
$\Delta,T\ll \mu$) it is straight-forward to derive \Eq{sumrule},
where
\begin{equation}
\rho_n(r) = \rho(r)\int d\xi \Big(-\frac{df(E)}{dE}\Big)_{E =
  \sqrt{\xi^2+\Delta^2(r)}}
\end{equation}
is the density of the normal-fluid component within the two-fluid
model.

Note that the temperature dependence of \Eq{sumrule} is different from
that of the number of normal particles, since in \Eq{sumrule} the
$r^6$ factor weights very strongly the surface of the system, where
$\Delta(r)$ is smaller and where consequently the normal-fluid
fraction $\rho_n/\rho$ is higher than in the center of the trap.
\end{appendix}


\begin{thebibliography}{*}
\bibitem{Anderson} M.H. Anderson, J.R. Ensher, M.R. Matthews,
  C.E. Wieman, and E.A. Cornell, Science \textbf{269}, 198 (1995);
  K.B. Davis, M.-O. Mewes, M.R. Andrews, N.J.
  van Druten, D.S. Durfee, D.M. Kurn, and W. Ketterle, Phys.
  Rev. Lett. \textbf{75}, 3969 (1995);
  C.C. Bradley, C.A. Sackett, J.J. Tollett, and
  R.G. Hulet, Phys. Rev. Lett. \textbf{75}, 1687 (1995).
\bibitem{DeMarco} B. De Marco and D.S. Jin, Science \textbf{285}, 1703
  (1999); B. De Marco, S.B. Papp, and D.S. Jin,
  Phys. Rev. Lett. \textbf{86}, 5409 (2001);
  A.G. Truscott, K.E. Strecker, W.I. McAlexander,
  G.B. Partridge, and R.G. Hulet, Science \textbf{291}, 2570 (2001);
  F. Schreck, G. Ferrari, K.L. Corwin, J. Cubizolles,
  L. Khaykovich, M.-O. Mewes, and C. Salomon, Phys. Rev. A \textbf{64},
  011402(R).
\bibitem{OHara} K.M. O'Hara, S.L. Hemmer, M.E. Gehm, S.R. Granade, and
  J.E. Thomas, Science \textbf{298}, 2179 (2002).
\bibitem{Houbiers} M. Houbiers, R. Ferwerda, H.T.C. Stoof,
  W.I. McAlexander, C.A. Sackett, and R.G. Hulet, Phys. Rev. A
  \textbf{56}, 4864 (1997).
\bibitem{BaranovPetrov1} M. A. Baranov and D. S. Petrov, Phys. Rev. A
  \textbf{58}, R801 (1998).
\bibitem{BruunCastin} G. Bruun, Y. Castin, R. Dum, and K. Burnett,
  Eur. Phys. J. D \textbf{7}, 433 (1999).
\bibitem{BaranovPetrov2} M.A. Baranov and D.S. Petrov, Phys. Rev. A
  \textbf{62}, 041601(R) (2000).
\bibitem{BruunMottelson} G.M. Bruun and B.R. Mottelson, Phys. Rev.
  Lett. \textbf{87}, 270403 (2001).
\bibitem{Kinast} J. Kinast, S.L. Hemmer, M.E. Gehm, A. Turlapov, and
  J.E. Thomas, Phys. Rev. Lett. \textbf{92}, 150402 (2004);
  M. Bartenstein, A. Altmeyer, S. Riedl, S. Jochim, C. Chin, J. Hecker
  Denschlag, and R. Grimm, Phys. Rev. Lett. \textbf{92}, 203201 (2004).
\bibitem{Roberts} J.L. Roberts, N.R. Claussen, S.L. Cornish,
  E.A. Donley, E.A. Cornell, and C.E. Wieman, Phys. Rev. Lett.
  \textbf{86}, 4211 (2000).
\bibitem{BaymPethick} G. Baym and C. Pethick, \textit{The Physics of
  Liquid and Solid Helium} (Wiley, New York, 1978), Part II, Chapter 1.
\bibitem{Lamb} H. Lamb, Proc. London Math. Soc. \textbf{13}, 189 (1882);
 G.F. Bertsch, Ann. Phys. (Leipzig) \textbf{86}, 138
  (1979); G.F. Bertsch and R.A. Broglia, \textit{Oscillations in Finite
  Quantum Systems} (Cambridge University Press, Cambridge, 1994).
\bibitem{Holzwarth} G. Holzwarth and G. Eckart, Z. Physik A
  \textbf{283}, 219-220 (1977).
\bibitem{Schwesinger} B. Schwesinger, Phys. Rev. C \textbf{29}, 1475
  (1984).
\bibitem{Ponomarev} V.Yu. Ponomarev, J. Phys. G \textbf{10},
  \textbf{L177} (1984).
\bibitem{Ca48Zr90} P. von Neumann-Cosel, F. Neumeyer, S. Nishizaki,
  V.Yu. Ponomarev, C.  Rangacharyulu, B. Reitz, A. Richter,
  G. Schrieder, D.I. Sober, T. Waindzoch, and J. Wambach,
  Phys. Rev. Lett. \textbf{82}, 1105 (1999)
\bibitem{Ni58} B. Reitz, A.M. van den Berg, D. Frekers, F. Hofmann,
  M. de Huu, Y. Kalmykov, H. Lenske, P. von Neumann-Cosel,
  V.Yu. Ponomarev, S. Rakers, A. Richter, G. Schrieder, K. Schweda,
  J. Wambach, and H.J. W\"ortche, Phys. Lett. B \textbf{532}, 179
  (2002).
\bibitem{Nesterenko} V.O. Nesterenko, J.R. Marinelli, F.F. de Souza
  Cruz, W. Kleinig, and P.-G. Reinhard, Phys. Rev. Lett. \textbf{85}, 3141
  (2000).
\bibitem{VinasRoth} X. Vi\~nas, R. Roth, P. Schuck, and J. Wambach,
  Phys. Rev. A \textbf{64}, 055601 (2001).
\bibitem{SchwesingerPingel} B. Schwesinger, K. Pingel, and G. Holzwarth,
  Nucl. Phys. A \textbf{341}, 1 (1988).
\bibitem{RingSchuck} P. Ring and P. Schuck, \textit{The Nuclear Many-Body
  Problem} (Springer-Verlag, Berlin, 1980).
\bibitem{deGennes} P.-G. de Gennes, \textit{Superconductivity of Metals and 
  Alloys}
(Benjamin, New York, 1966).
\bibitem{Bulgac} A. Bulgac and Y. Yu, Phys. Rev. Lett. \textbf{88},
  042504 (2002). 
\bibitem{Grasso} M. Grasso and M. Urban, Phys. Rev. A \textbf{68}, 033610
  (2003).
\bibitem{Abraham} E.R.I. Abraham, W.I. McAlexander, J.M. Gerton,
  R.G. Hulet, R. Cot\'e, and A. Dalgarno, Phys. Rev. A \textbf{55},
  R3299 (1997).
\bibitem{Farine} M. Farine, P. Schuck, and X. Vi{\~n}as, Phys. Rev. A
  \textbf{62}, 013608 (2000).
\bibitem{UrbanSchuck} M. Urban and P. Schuck, Phys. Rev. A
  \textbf{67}, 033611 (2003).
\bibitem{Leggett} A. J. Leggett, Phys. Rev. \textbf{140}, A1869 (1965);
  Phys. Rev. \textbf{147}, 119 (1966).
\bibitem{Betbeder} O. Betbeder-Matibet and P. Nozi\`eres,
  Ann. Phys. (N.Y.) \textbf{51}, 392 (1969).
\end{thebibliography}
\end{document}